# Understanding The Impact Of Socialbot Attacks In Online Social Networks

**Silvia Mitter**
Knowledge Technologies Institute
Graz University of Technology, Austria
smitter@student.tugraz.at

**Claudia Wagner**
Institute for Information and Communication Technologies
JOANNEUM RESEARCH Graz, Austria
claudia.wagner@joanneum.at

**Markus Strohmaier**
Knowledge Technologies Institute
Graz University of Technology, Austria
markus.strohmaier@tugraz.at

## Abstract
Online social networks (OSN) like Twitter or Facebook are popular and powerful since they allow reaching millions of users online. They are also a popular target for socialbot attacks. Without a deep understanding of the impact of such attacks, the potential of online social networks as an instrument for facilitating discourse or democratic processes is in jeopardy. In this extended abstract we present insights from a live lab experiment in which social bots aimed at manipulating the social graph of an online social network, in our case Twitter. We explored the link creation behavior between targeted human users and our results suggest that socialbots may indeed have the ability to shape and influence the social graph in online social networks. However, our results also show that external factors may play an important role in the creation of social links in OSNs.

## ACM Classification Keywords
K.4.2 Computing Milieux - Computers and Society: Social Issues

## Introduction

Online social networks (OSNs) like Facebook and Twitter can be used to spread misinformation and propaganda, as one could for example see during the US political elections [4]. Recently *socialbots*, which are automated agents in OSNs that can perform certain actions on their own, have spread in OSNs. Past research highlights the dangers of socialbots [2] and shows that Facebook can be infiltrated by sending automated friend requests to users. The average acceptance rate of such automated friend requests was 36.7% which could be as high as 80% when common friends were present. Other research [5] shows that users susceptible to socialbot attacks may exhibit certain characteristics that allow us to distinguish them from non-susceptible users.

In this work we focus on social bot attacks and aim to enhance our understanding of the impact of socialbot attacks by presenting an empirical study on the ability of socialbots to shape or influence the social graph of Twitter. We address the following research question:

*Can socialbots be used to influence link creation between targeted human users in OSNs?*

We analyze data from a socialbot experiment on Twitter which was conducted by the *Pacific Social Architecting Corporation* (PacSocial) in 2011 [3]. The aim of the experiment was to explore the impact of socialbots on creating links between users (*targets*). In our study we investigate to what extent socialbots were able to shape the social graph by analyzing the impact the socialbots had on the link creation behavior of the targets in more detail as this was done before. We found that a large proportion of social links created in OSNs cannot be explained solely by the observational data from the OSN. Our results suggest that also external factors (which may vary over time) drive link creation behavior and those factors may function as confounding variables for the results of similar studies, such as [1].

## Experimental Setup

In the following section we describe the design of our empirical study and introduce the dataset. Then, we present measures which allow us to assess socialbot impact and success to explore whether socialbots can be used to influence link creation between users in OSNs.

### Dataset

The dataset was provided by PacSocial, and was collected during a socialbot competition in 2011. Additionally tweets from targeted users and the bots were collected by us. The experiment was designed to consist of a *control phase* (*ctr*) of 33 days and an *experimental phase* of 21 days. The socialbots were launched immediately after the control phase. The main objective of the socialbots was to cause the creation of new social links between users within a selected target group. Another objective was to simply interact with the targets. Nine socialbots were launched and every socialbot had its own target group each consisting of 300 user accounts. The purpose of the control phase was to observe the link creation behavior of the target users under normal conditions as a baseline information. The authors of this paper did not participate in the design, setup or execution of the competition in any way.

| | |
|---|---:|
| Number of Socialbots | 9 |
| Number of Targets | 2,700 |
| Number of Targets following Socialbots | 192 |
| Number of Targets communicating with Socialbots | 232 |
| Number of Tweets | 1,006,351 |

*Table 1: Dataset description*

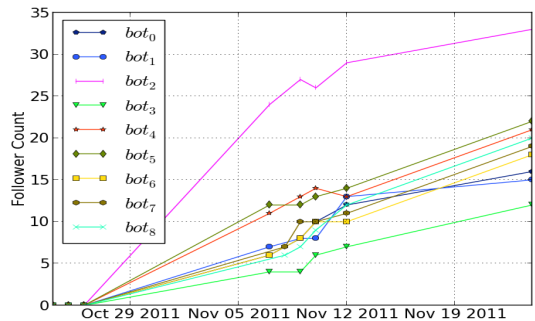

*Figure 1: Number of targeted users following the socialbots, shown per socialbot and its target groups each consisting of 300 users .*

Table 1 provides an overview of the dataset, which consists of tweets that were published by any of the socialbots or target users during the control and experimental phase, as well as tweets from the targets (3,200 at maximum), starting approximately two months prior to the control phase. The additional data from the dataset was collected via the Twitter API. Figure 1 shows how many of the targeted users were following the socialbots, where values captured at specific moments in time are indicated by markers.

Figure 2 gives an overview of the number of tweets authored by socialbots where they were recommending users to each other. We manually inspected a sample of those recommendation tweets authored by socialbots and observed that they most commonly either address one user and recommend one other user (e.g., @UserA - you would like my #friend @UserB), or recommend several users within one tweet. We are not aware of how the socialbots chose which users to recommend to each other.

Based on the fact that socialbots still were active after the end of the original experimental phase, we decided to pick a modified *experimental phase 2* (exp2), which lasted until the last available follow information of socialbots and targets in the dataset. Figure 3 provides an overview of the control phase, experimental phase 1 and experimental phase 2.

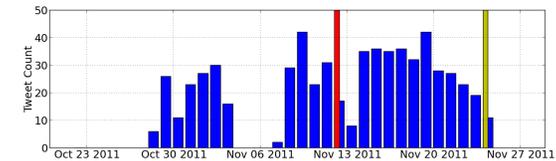

*Figure 2: Number of tweets where socialbots recommend users to each other. The first (red) line indicates the end of experimental phase 1, the second (yellow) line indicates the end of experimental phase 2 (see Figure 3 for more details about the experimental phases).*

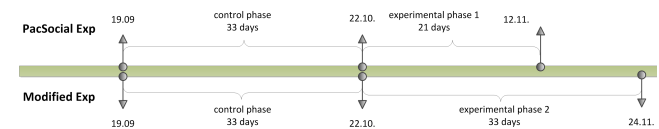

*Figure 3: Overview of control phase, experimental phase 1 (original phase) and experimental phase 2 (modified phase)*

*Measuring Socialbot Impact and Success*
A previous study on this dataset [3] found a significant increase (of approximately 43%) in the link creation activities of target users within the target groups during the experimental phase 1 compared to the control phase, which led to the assumption that the socialbots were very successful in creating new links between targets. However, the authors did not further explore if other factors may have caused the link creation. In our work we aim to address this open issue. Therefore, we introduce several *success measures,* which can be seen as preceding situations potentially causing the link creation between two targeted users. The success measures are described on two different dimensions: The first dimension defines *who* may cause the link creation by defining several mediator types. The second dimension describes *how* link creation may be motivated by defining recommendation types.

MEDIATORS

If two users did not have any direct interactions in the past a third-party *mediator* may have caused a link creation. We distinguish between following mediators:

- *Human Mediator*: May be any user account in the target group.
- *Socialbot Mediator:* A link is created between two users after a socialbot mediator, but no human mediator, was observed.
- *Human-&-Socialbot Mediator:* Human and socialbot mediators were observed before link creation.
- *No Measurable Mediator:* No potential mediator can be identified. This category for example captures the fact that the link creation in OSNs can exclusively be motivated by real life factors which are not reflected in the dataset captured by the OSNs.

RECOMMENDATION TYPES (RT)

Recommendation types which may cause new links between users are defined as shown in Figure 4(a-c).

For links established by direct user interaction (see Figure 4 (d)) we do not assume that any mediator and recommendation type was involved in the link creation. We compute our impact measures by combining recommendation types with different types of mediators. Therefore, our measures which capture all potential combinations of recommendation types and involved mediators explain a large variety of possible causes of link creation. Potential recommendations created by Twitter itself cannot be considered since we do not have any information about them.

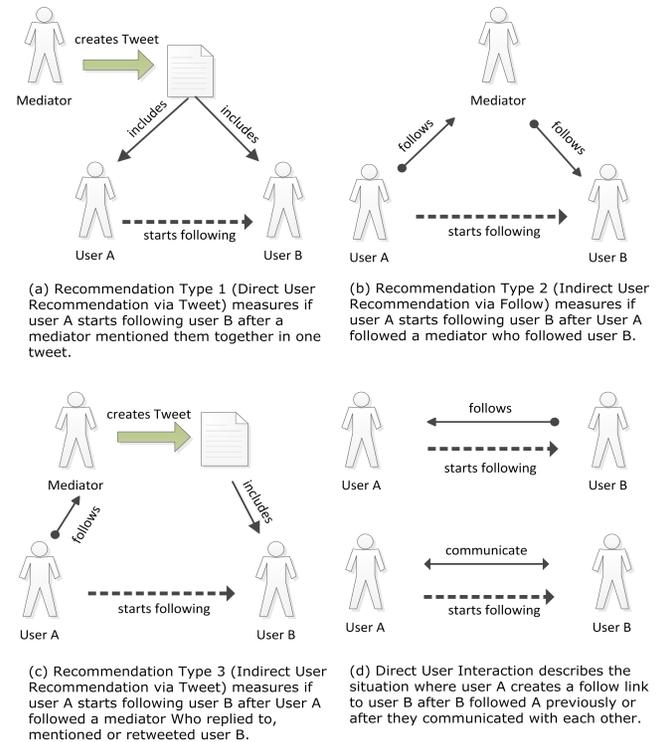

(a) Recommendation Type 1 (Direct User Recommendation via Tweet) measures if user A starts following user B after a mediator mentioned them together in one tweet.

(b) Recommendation Type 2 (Indirect User Recommendation via Follow) measures if user A starts following user B after User A followed a mediator who followed user B.

(c) Recommendation Type 3 (Indirect User Recommendation via Tweet) measures if user A starts following user B after User A followed a mediator Who replied to, mentioned or retweeted user B.

(d) Direct User Interaction describes the situation where user A creates a follow link to user B after B followed A previously or after they communicated with each other.

*Figure 4: Recommendation Types (a-c) and Direct User Interaction (d).*

**RESULTS**

Our results vary for different success measures as expected. The success measures allow differentiating between links which might have been created anyway since the users were already indirectly related before but had no direct connection (*human mediated*), links which were most likely caused by socialbot mediators (*socialbot mediated*) and links which were most likely caused by socialbot *or* human mediators (*human-&-socialbot mediated*). Table 2 shows the basis for applying our success measures. The total number of created links is reduced by the number of links created with preceding direct user interaction (Figure 4 (d)), which indicates that those links are created based on the fact that the users already knew each other. Values are summed over all 9 target groups and averaged per day.

*Table 2:* Link creation summed over target groups averaged per day.

|                          |       |       |       |
|--------------------------|-------|-------|-------|
| Total                    | 5.49  | 7.62  | 6.76  |
| -Direct User Interaction | -2.12 | -2.71 | -2.55 |
| **Basis**                | **3.36** | **4.91** | **4.21** |

In Table 3 we show the proportion of newly created links (corresponds to the row *Basis* from Table 2) for all three recommendation types combined (RT123), split by the different mediator types, for the control phase (*ctr*), experimental phase 1 (*exp1*) and experimental phase 2 (*exp2*).

| Link Creation RT 123 | human mediated | | socialbot mediated | | human & socialbot mediated | | undefined mediated | |
|---|---|---|---|---|---|---|---|---|
| | abs | % | abs | % | abs | % | abs | % |
| control phase | 1.49 | 44.14 | 0.00 | 0.00 | 0.00 | 0.00 | 1.88 | 55.86 |
| exp. phase 1 | 1.81 | 36.90 | 0.33 | 6.79 | 0.29 | 5.83 | 2.48 | 50.48 |
| exp. Phase 2 | 1.46 | 34.54 | 0.49 | 11.51 | 0.49 | 11.51 | 1.79 | 42.45 |

*Table 3: Newly created links (corresponds to the row Basis from Table 2) split by mediator types. Values are shown for control phase (ctr), experimental phase 1 (exp1) and experimental phase 2 (exp2) for the combination of all recommendation types (RT 123).*

Table 3 shows the following results for the *experimental phase 1*: The success of socialbots for RT 123 (which is the combination of all three recommendation types, showed in Figure 4) was only 6.79%, although the overall number of link creation increased significantly. Also the overlap of potentially bot and human based link creation is rather small (5.83%). A significant part of links was created after a preceding human mediator (36.90%) recommendation could be measured and the majority of new links were created without a measurable mediator (50.48%). This means that no potential cause can be observed from the data. This is not surprising since also real world factors may impact the creation of social links and therefore function as mediating events. In summary, our results show that the observable impact of socialbots is rather low in experimental phase 1, while the total increase of links was rather high.

Interestingly, for the *experimental phase 2* which measured consequences of bot activity a bit longer, higher socialbot success could be observed: Although the overall number of link creation increases less during *exp2* than during *exp1*, we can identify a larger proportion of new links which *are possibly caused by socialbots* (11.51% for RT-123), and also a much larger proportion of newly created links where socialbot and human preceding recommendation could have been measured (11.51% for RT-123), while the proportion of potentially human based link creation is approximately the same as for experimental phase 1 (34.54% for RT-123). However, one needs to note that we also found that for a large proportion (42.45% for RT-123) of newly created links no explanatory causes could be identified from the data. This means that for RT-123 about 23% of newly created links are potentially based on socialbots or on humans and socialbots.

Comparing results from both studies shows an increase in link creation in both experiments (see Table 3). However, we found *no evidence* that the dramatic increase for experimental phase-1 was caused by socialbots. Our results suggest that the proportion of links caused by social bots may be increased if attacks are conducted over a longer duration. However, our results also show that in both experimental phases factors outside the dataset, such as real world events and factors outside the OSN, may play an important role. Our findings are partly in line with previous studies on predicting social links in OSNs which also show that external factors may impact the link creation behavior of users. The approach used in [1] allows recommending new social links to active Facebook users with high precision. They show that out of 20 friendships they recommended nearly 40% of them were realized in the near future.

## Conclusions
In this work we report results from a live lab experiment on Twitter in which social bots targeted users of an OSN. While our results suggest that socialbots indeed can play a role in changing the fabric of OSN, our results also highlight the role of external factors in link creation. Specifically, our results show that there is not necessarily a direct causal relation between the increase of the number of links between targets and the socialbots interactions and that further research is required to explore the impact of external factors (e.g., offline events). We hope that our work represents a stepping stone for more principled investigations into the role of socialbots in OSNs such as Twitter or Facebook. We believe that a better understanding of such attacks is essential in ensuring that OSNs become a trustworthy and effective tool for exchanging of ideas and information.

## Acknowledgements
We want to thank the members of the PacSocial, especially Tim Hwang for sharing the dataset as well as Ian Pierce and Max Nanis for technical support. This work was supported in part by a DOC-fForte fellowship of the Austrian Academy of Science to Claudia Wagner.